\documentclass[prl,showpacs,preprintnumbers]{article}
\usepackage{amsmath,bbm,amssymb}
\usepackage{dcolumn}
\usepackage{graphicx}
\usepackage{makeidx}
\usepackage{bm}

\author{Clovis Jacinto de Matos\footnote{ESA-HQ, European Space Agency, 8-10 rue Mario Nikis,
75015 Paris, France, e-mail: Clovis.de.Matos@esa.int}}

\title{Are Superfluid Vortices in Pulsars Violating the Weak Equivalence Principle?}

\begin{document}

\maketitle \begin{abstract} In the present paper we argue that
timing irregularities in pulsars, like glitches and timing noise,
could be associated with the violation of the weak equivalence
principle for vortices in the superfluid core of rotating neutron
stars.
\end{abstract}

\maketitle

\section{Introduction}

Pulsars are traditionally used to test general relativity
\cite{Taylor}. Anomalous inertial mass excess have been found for
Copper pairs in rotating supercondutors \cite{Tate01}
\cite{Tate02} and for superfluid vortices in rotating superfluids
\cite{Duan}. These anomalies can be understood in terms of a
breaking of the weak equivalence principle for Cooper pairs in
superconductors \cite{de Matos1} and for vortices in superfluids
\cite{de Matos2}. Can we find phenomenological evidence of similar
effects in the context of pulsar physics? In the present paper we
argue that the answer to this question is positive. Glitches and
timing noise of pulsars \cite{Lyne} might be attributed to a
violation of the weak equivalence principle for the superfluid
neutron vortices. Although the magnitude of this effect is below
the maximum threshold imposed by current tests of the strong
equivalence principle in binary systems containing at least one
pulsar, it is shown to be too small to be detected by current
observational instrumentation.

Section 2 contains a review of the current tests of the
equivalence principle in classical and quantum physics (particle
physics and superconductors) in the Earth laboratory, and with
pulsars. Section 3, discusses the subject of the equivalence
principle for superfluids. Section 4 introduces the subject of
superfluid physics in the description of the dynamics of pulsars,
and reports about the measurement and current physical
understanding of glitches and timing noise. In section 5 a
possible interpretation of glitches in terms of a breaking of the
weak equivalence principle for neutron vortex lines within pulsars
is proposed. In section 6 this theoretical model is compared with
Packard metastable vortex model to account for similar phenomena
in Pulsars. In conclusion the parallels which can be established
between anomalous mass measurements in superconductors,
superfluids and neutron star physics, are critically assessed.

\section{Testing the Equivalence Principle on Earth and in Pulsars}

The Strong Equivalence Principle (SEP) is completely embodied into
general relativity, while alternative theories of gravity predict
a violation of some or all aspects of SEP. The SEP is, according
to its name, stronger than both the Weak Equivalence Principle
(WEP) and the Einstein Equivalence Principle (EEP). The WEP states
that all test bodies in an external gravitational field experience
the same acceleration regardless of the mass and composition.
While the WEP is included in all metric theories of gravity, the
EEP goes one step further and also postulates Lorentz-invariance
and positional invariance. Lorentz-invariance means that no
preferred frame exists, so the outcome of a local
non-gravitational experiment is independent from the velocity of
the apparatus, while positional invariance renders it unimportant
where this experiment is being performed. The SEP includes both
the WEP and the EEP, but postulates them also for gravitational
experiments.

The violation of WEP means that the ratio of the gravitational
mass, $m_g$, and the inertial mass, $m_i$, of two bodies $A$ and
$B$ falling freely, under the single influence of a homogeneous
gravitational field, are not equal to each other. This is usually
quantified through the E\"{o}tv\"{o}s-factor, $\eta(A,B)$.
\begin{equation}
\eta(A,B)= \Big(\frac{m_g}{m_i}\Big)_A -
\Big(\frac{m_g}{m_i}\Big)_B\label{w1}
\end{equation}
The E\"{o}tv\"{o}s-factor is usually obtained from the measurement
of the differential acceleration, $\Delta a$, of two test bodies,
$A$ and $B$, falling freely in the gravitational field $g$.
\begin{equation}
\eta(A,B)=\frac{\Delta a}{g} \label{w2}
\end{equation}
The E\"{o}tv\"{o}s-factor, can also be estimated from the
measurement of the relative differential rotational frequency,
$\Delta \omega$, of two test bodies, $A$ and $B$, freely rotating
in a gravitomagnetic field $B_g$.
\begin{equation}
\eta(A,B)=\frac{\Delta \omega}{B_g} \label{w2a}
\end{equation}
The Newtonian gravitational field $g$, expressed in the SI unit
system as an acceleration in  $m/s^2$, and the gravitomagnetic
field, with SI units of angular velocity $Rad/s$, appear both in
the weak field linear approximation of Einstein field equations.
In this theoretical framework acceleration fields from
gravitational and non-gravitational origin cannot be physically
distinguished from each other, the same restriction applies also
for angular velocities from gravitomagnetic and
non-gravitomagnetic origin. As argued by Anandan \cite{Anandan},
any form of the principle of equivalence cannot be demonstrated on
a purely theoretical basis. Thus it can only be justified by
experiment.

Current experimental measurements \cite{Bassler} \cite{Smith}
indicate that the WEP is verified, for classical macroscopic
systems, i.e. systems which do not break gauge invariance, with a
fractional precision of the E\"{o}tv\"{o}s-factor
\begin{equation}
\eta < 5 \times 10^{-13}\label{e1}
\end{equation}

Contrasting with classical physics, in quantum mechanics the
motion of a particle in the presence of an external gravitational
field is mass dependent \cite{Stella}. Collela, Overhauser and
Werner, (COW) \cite{cow} measured the phase shift $\delta \phi$
induced by gravity on a monoenergetic neutron beam propagating
with velocity $v_{n_0}$ in the arms of a Fabry-Perot
interferometer located in two different gravitational potentials.
\begin{equation}
\delta \phi=m_{n^0}^2 g l_1 l_2 \lambda / \hbar^2
\end{equation}
where $l_1$ and $l_2$ are the length of the interferometer arms
located in two different gravitational potentials, and the height
separating the two arms respectively, $\lambda=\hbar/m_{n^0}
v_{n^0}$ is the de Broglie wavelength of the cold neutron beam,
and $m_{n^0}$ is the neutron mass. COW experiments revealed that:
\begin{enumerate}

\item The phase shift due to gravity is seen to be verified to well
within $1\%$.

\item The gravitational Newtonian potential enters into the
Schrodinger equation as expected.

\item Gravity is not purely geometric at quantum level because the
effect depends on $(m_{n^0}/\hbar)^2$.

\end{enumerate}
Making the distinction between the neutron gravitational mass
$m_{n^0g}$ and the neutron inertial mass $m_{n^0i}$,
$(m_{n^0}/\hbar)^2$ should be replaced by $m_{n^0i}
m_{n^0g}/\hbar^2$. Thus COW experiments indicate that $m_{n^0g} =
m_{n^0i}$ within $1\%$ for neutrons.

Still in the domain of quantum mechanics, but at macroscopic
scales: For superconductors, which do break gauge invariance, the
equivalence principle has also been tested to some extent. Cabrera
and Tate \cite{Tate01, Tate02}, through the measurement of the
magnetic trapped flux originated by the London moment, reported an
anomalous Cooper pair inertial mass excess in thin rotating
Niobium superconductive rings:
\begin{equation}
\Delta m_i=m_i^*-m_i=94.147240(21)eV\label{e3}
\end{equation}
Here $m_i^*=1.000084(21)\times 2m_e=1.023426(21)MeV$ ($m_e$ being
the standard electron mass) is the experimentally measured Cooper
pair inertial mass (with an accuracy of 21 ppm), and
$m_i=0.999992\times2m_e=1.002331 MeV$ is the theoretically
expected Cooper pair inertial mass including relativistic
corrections.

This anomalous Cooper pair mass excess has not received, so far, a
satisfactory explanation in the framework of superconductor's
physics. If the gravitational mass of the Cooper pairs, $m_g$,
remains equal to the expected theoretical Cooper pair inertial
mass, $m_i=m_g=0.999992\times2m_e=1.002331 MeV$, Tate's experiment
would reveal that the Cooper pairs break the WEP with an
E\"{o}tv\"{o}s-factor, $\eta(E,T)=9.19\times 10^{-5}>>5\times
10^{-13}$, obtained from eq.(\ref{w1}) assuming the Experimental
$(E)$ and Theoretical $(T)$ ratios $m_g/m_i^*=0.999908$ and
$m_g/m_i=1$ respectively. The question is thus: Is an excess of
mass, similar to the one observed by Tate for the cooper pairs
inertial mass, also occurring for the cooper pair's gravitational
mass? In recent work with Christian Beck the author derived a law
for the breaking of the WEP for Cooper pairs based on an
electromagnetic model of dark energy for the bosonic vacuum
fluctuations in superconductors \cite{de Matos1}.
\begin{equation}
\eta\sim\frac{3\ln^4 3}{8 \pi}\frac{k^4G}{c^7\hbar^3 \Lambda}
T_c^4\label{12}.
\end{equation}
Remarkably, this equation connects the five fundamental constants
of nature $k,G,c,\hbar, \Lambda$ with measurable quantities in a
superconductor, $\eta$ and $T_c$.

In 1987 Jain et al carried out an experiment to probe the SEP for
Cooper pairs \cite{Jain}. The experiment consisted of two
Josephson junctions located in the Earth gravitational field at
different heights, connected in opposition by superconducting
wires. Jain et al. experiment has shown that using the Cooper
pairs as probe masses, we also reach the conclusion that the
laboratory is accelerating with respect to a local Minkowski
spacetime. This plainly justifies the curved spacetime
description, which has been well tested for classical matter, to
hold for Cooper pairs as well. This experiment also demonstrated
that the inertial and gravitational mass of Cooper pairs are
exactly equal to each other within an accuracy of $4\%$:
\begin{equation}
\frac{m_i}{m_g}=1\pm0.04
\end{equation}
Unfortunately the accuracy of Jain's experiment is not good enough
to discard or confirm a difference between the inertial and the
gravitational mass of Cooper pairs of 21 ppm streamlined with Tate
et al experiments. Tajmar et al. carried out experiments in 2009
to test the WEP for high Tc superconductors using a magnetic
suspension balance \cite{Tajmar}. although this experiment
achieved an higher accuracy than Jain's experiment, it could only
resolve E\"{o}tv\"{o}s-factor, $\eta< 2\times 10^{-3}$, being 2
orders of magnitude away from the E\"{o}tv\"{o}s-factor
$\eta(E,T)=9.19\times 10^{-5}$ calculated from Tate's experiment.

A violation of SEP means that objects with different fractional
mass contributions from self-gravitation origin would fall
differently in an external gravitational field. This is quantified
by the parameter $\zeta$, defined through
\begin{equation}
\Big(\frac{m_g}{m_i}\Big)_A=1+\zeta \Big(\frac{E_g}{mc^2}\Big)_A
\label{w3}
\end{equation}
where $m_g$, $m_i$, $E_g$ are respectively the gravitational mass,
inertial mass, and gravitational self-energy of body $A$. For a
binary system composed of two bodies $A$ and $B$, in free fall in
a gravitational field $g$ while orbiting around each other, a non
zero value of $\zeta$ would result in the polarization of the
binary orbits in the direction of $g$ \cite{Nordtvedt1}
\cite{Nordtvedt2}, resulting in a (small) forced eccentricity of
the orbit. In this type of systems the parameter to be constrained
is $\Delta$, it is similar to $\zeta$ but without the requirement
of linear dependence on the self-gravitational energy, $E_g/mc^2$.
It is defined for an individual body $A$ by
\begin{equation}
\Big(\frac{m_g}{m_i}\Big)_A=1+\Delta_A\label{w4}
\end{equation}
Dynamics of a binary orbit depends on the difference
\begin{equation}
\eta=\Delta_A-\Delta_B\label{w5}
\end{equation}
between the two objects $A$ and $B$ \cite{Damour}, where $\eta$ is
given by eq.{\ref{w1}). Of course, this effect is most detectable
in binary systems composed by bodies with radically different
gravitational self energies like, for example, a pulsar orbiting a
white dwarf.

Pulsars are rotating neutron stars. The signals received from the
pulsars are modulated at a remarkably constant frequency which is
resulting from the rotation frequency of the star. The orbital
parameters of the binary systems, containing at least one pulsar,
are all deduced by fitting the arrival times of pulses. Pulsars
are well established test beds for relativity. Observation of the
neutron star-neutron star binary PSR B1913+16 have established
that its orbit decays at the rate predicted by general relativity
within 0.3$\%$ \cite{Taylor}. The observation by Wex of an
ensemble of long orbit pulsars yields a limit of $|\eta|< 0.009$
at $95\%$ confidence level \cite{Wex}, which represents the most
stringent test of SEP violation until the present date. As we will
see later, this result is also useful to assess the validity of
the WEP in superfluids, since the neutron star core is a
superfluid.

In the context of experiments in the Earth laboratory, the WEP is
poorly tested for superfluids in general, and superfluid vortices
in particular. However the vortex inertial mass in superfluid
Helium has been extensively discussed in the literature
\cite{Duan}\cite{Thouless}, \cite{Popov}.

\section{Weak Equivalence Principle in Superfluids}

A vortex line in rotating superfluid Helium 4 is a topological
singularity, which consists of a normal core region of the size of
the coherence length $\xi$, and an outside region of circulating
supercurrent. The coherence length can be estimated from the
Heisenberg uncertainty principle.
\begin{equation}
\xi\sim\frac{\hbar}{m c_s}\label{v1}
\end{equation}
where $m$ is the bare atomic mass in $^4 He$ and $c_s$ is the
speed of sound in the superfluid. Taking $c_s\sim 2\times 10 ^2
m/s$ \cite{Nozieres} we estimate $\xi\sim 1 {\AA}$. In the
theoretical framework of the classical fluid model the only
obvious contribution to the vortex mass is the core
mass\cite{Baym}.
\begin{equation}
m_{core}=L \pi \xi^2 \rho\label{v2}
\end{equation}
where $L$ is the length of the vortex line, and $\rho=Nm$ is the
density with $N$ the bulk number density of $^4 He$ atoms. This
small vortex mass is usually discarded in the equations of motion
of vortex dynamics since it contradicts experimental data.

Duan in \cite{Duan} shown that due to spontaneous gauge symmetry
breaking in superfluids, the condensate compressibility
contributes to a vortex mass which is much larger than the
classical core mass. He calculates that the vortex inertial mass
turns out to diverge logarithmically with the system size.
\begin{equation}
m_{inertial}=m_{core}\ln \Big( \frac{L}{\xi} \Big)\label{v3}
\end{equation}
Where $L$ is the length of the vortex. For a practical superfluid
system in the Earth laboratory, $\ln(L/\xi)\sim20-30$.

The number of vortices $N_v$ appearing in a cylindrical sample of
superfluid $^4 He$ rotating with angular velocity $\Omega$ is
deduced from the quantization of the vortex canonical momentum.
\begin{equation}
N_v=\frac{2\pi R^2 \Omega}{\hbar/m}\label{v4}
\end{equation}
where $R$ is the radius of the superfluid sample. The total
increase of the inertial mass of a rotating superfluid sample with
respect to the same non-rotating sample, is obtained from
eq.(\ref{v3}) and eq.(\ref{v4})
\begin{equation}
\Delta M_{inertial}=N_v m_{core}\Big( \ln\Big(\frac{L}{\xi}\Big)
-1\Big)\label{v5}
\end{equation}
Assuming that the weak equivalence principle is still valid in
superfluids this overall increase of inertial mass should appear
together with a similar increase of the gravitational mass of the
superfluid sample.
\begin{equation}
\Delta M_{inertial}=\Delta M_{gravitational}\label{v6}
\end{equation}
Thus we should observe an increase of the weight of the rotating
superfluid sample with respect to the same sample in the
stationary state. Taking a cylindrical sample of radius $R=1 cm$
and $\ln(L/\xi)\sim20-30$, rotating at $\Omega=1 Rad/s$ in
eq.(\ref{v5}), we estimate that the total increase of
gravitational mass is of the order of $\Delta
M_{gravitational}=10^{-14}-10^{-9} Kg$. Thus the experimental
detection of the associated increase of weight of the overall
sample is a challenging task to perform, that has not yet been
overcome by experimentalists in the Earth laboratory. In summary
Until the present date the weak equivalence principle has not been
tested for superfluid vortices.

The breaking of gauge symmetry makes the superfluid sample a
preferred frame, this should be associated with a speed of light
in the superfluid vacuum different from its classical value $c_0$,
appearing in Lorentz transformations. As demonstrated by Duan and
Popov \cite{Duan} \cite{Popov} the vortex inertial mass can be
expressed in function of the vortex static energy $\epsilon_0$
which is also logarithmically divergent as the sample size.
\begin{equation}
m_{inertial}=\frac{\epsilon_0}{c_s^2}\label{v8}
\end{equation}
where $c_s$ is the speed of sound in the superfluid.

Starting from Mach's principle, which asserts that there is a
connection between the local laws of physics and the large scale
properties of the universe, Sciama in \cite{Sciama} introduced the
relation
\begin{equation}
c_0^2=\frac{2GM}{R}\label{m1}
\end{equation}
where $R$ and $M$ are the radius and the baryonic mass of the
universe. Einstein's relationship linking energy and mass then
takes the form
\begin{equation}
E=mc_0^2=\frac{2GMm}{R}\label{m2}
\end{equation}
which can be interpreted as a statement that the inertial energy
that is present in any physical object is due to the gravitational
potential energy of all the matter in the universe acting on the
object. Therefore the mass $m$ appearing in eq.(\ref{m2}) should
be the gravitational mass of the object.
\begin{equation}
E=m_{gravitational} ~c_0^2\label{m3}
\end{equation}
Since the rest mass energy of the vortex $\epsilon_0$ must be
conserved independently of the effective value of the vacuum speed
of light, the gravitational mass will adjust its value to
compensate the variation of the speed of light in the superfluid
vacuum.
\begin{equation}
m_{gravitational} ~c_0^2=m_{core}c_s^2
\ln\Big(\frac{L}{\xi}\Big)\label{m4}
\end{equation}
From eq.(\ref{m4}) we deduce that the gravitational mass of a
superfluid vortex $m_{gravitational}$ is proportional to the
classical vortex core mass and also diverges logarithmically as
the size of the vortex.
\begin{equation}
m_{gravitational}=\Big(\frac{c_s}{c_0} \Big)^2 m_{core}\ln \Big(
\frac{L}{\xi} \Big ) \label{v7}
\end{equation}
where the proportionality coefficient is equal to the square of
the ratio between the speed of sound in the superfluid $c_s$ and
the classical speed of light in vacuum $c_0$. Comparing
eq.(\ref{v3}) and eq.(\ref{v7}) we conclude that due to the
principle of energy conservation and to the breaking of gauge
invariance in superfluids the inertial and the gravitational mass
of a vortex cannot be equal to each other. Therefore the weak
equivalence principle should break for the case of superfluid
vortices.

As we have shown above, measuring the vortices gravitational mass
comparing the weight of the superfluid sample in rotating and
stationary state is challenging due to the extremely small value
of the vortex core mass. However in free fall experiments with
rotating superfluid samples it should be possible to measure the
differential acceleration $\Delta a$ between the vortex and the
bulk superfluid. The E\"{o}tv\"{o}s factor $\eta$ associated with
the free fall of a vortex and the superfluid bulk under the single
influence of the Earth gravitational field $g_0$ would be obtained
from eq.(\ref{w2}):
\begin{equation}
\eta=\frac{\Delta a}{g_0}\label{vvv1}
\end{equation}
Let us first assume that the friction force between the vortex and
the superfluid bulk is null. On one side, since the superfluid
bulk inertial and gravitational mass are equal, the center of mass
of the superfluid bulk will fall with and acceleration
\begin{equation}
a_{superfluid}=g_0\label{vv1}
\end{equation}
On the other side The vortex will fall according to the equation
of motion
\begin{equation}
g_0 \;  m_{gravitational}=m_{inertial} \; a_{vortex} \label{v9}
\end{equation}
substituting eq.(\ref{v7}), and eq.(\ref{v3}) in eq.(\ref{v9}) we
calculate the vortex falling acceleration
\begin{equation}
a_{vortex}=g_0 \frac{c_s}{c}\label{vv2}
\end{equation}
Substituting the accelerations $a_{superfluid}$, eq(\ref{vv1}),
and $a_{vortex}$, eq.(\ref{vv2}), in eq.(\ref{vvv1}) we obtain the
E\"{o}tv\"{o}s factor $\eta$ for a superfluid vortex with respect
to the superfluid bulk.
\begin{equation}
\eta=1-\Big(\frac{c_s}{c_0}\Big)^2 \label{v10}
\end{equation}
Taking $c_s\sim2\times 10^2 m/s$ we have $\eta\sim 1$ which is
much higher than the upper limit measured for classical material
systems of $5\times 10^{-13}$, eq.(\ref{e1}).

If instead of assuming no friction between the vortices and the
superfluid bulk, like we did above, we assume an ideal rigid
connection between both systems. We deduce from the equation of
motion of the freely falling rotating superfluid sample, a falling
acceleration $a_z$.
\begin{equation}
a_z=
\frac{1+\Big(\frac{c_s}{c}\Big)^2\frac{m_v}{m}}{1+\frac{m_v}{m}}g_0\label{v12}
\end{equation}
where $m$ is the total classical mass of the superfluid bulk
(without the vortices) and $m_v=N_v m_{core}
\ln\Big(\frac{L}{\xi}\Big)$ is the total inertial mass of vortices
in the superfluid sample, with $N_v$ being the effective number of
vortices. Comparing this acceleration with the falling
acceleration of the same non-rotating sample, $g_0$, we calculate
the E\"{o}tv\"{o}s factor $\eta'$ of the rotating sample with
respect to the non-rotating one.
\begin{equation}
\eta'=\frac{g_0 - a_z }{g_0} \label{v13}
\end{equation}
substituting eq.(\ref{v12}) in eq.(\ref{v13}) we obtain \cite{de
Matos2}
\begin{equation}
\eta'=\frac{m_v}{\Delta m}\eta\label{v14}
\end{equation}
where $\Delta m=m-m_v$ and $\eta=1-\Big(\frac{c_s}{c_0}\Big)^2$ is
the E\"{o}tv\"{o}s factor of one vortex with respect to the
superfluid bulk (assuming no friction between both),
eq.(\ref{v10}). Taking a cylindrical sample of radius $R=1 cm$ and
$\ln(L/\xi)\sim20-30$, rotating at $\Omega=1 Rad/s$ in
eq.(\ref{v14}), we estimate the order of magnitude of $\eta'\sim
10^{-11}$, which is 2 orders of magnitude above the upper limit
experimentally determined for normal materials, which do not break
gauge invariance, eq.(\ref{e1}).

\section{Glitches and Timing Noise in Pulsars}
Pulsars are rotating neutron stars, which consist of a solid iron
crust and a superfluid neutron core containing also a
superconducting proton layer (Ginzburg, 1971). Since the protons
represent only a few percent of the total star, we will neglect
them in the discussion which follows, and will concentrate on the
superfluid neutron part. The temperature of the star is probably
$\sim 10^8 K$, and the neutron superfluid phase density is
$~10^{17} Kg/m^3$, so that the neutrons are highly degenerate.
Migdal \cite{Migdal} was the first to propose that the neutrons
near the Fermi surface might be paired in such a way as to suffer
a BCS type condensation. According to Hoffberg et al.
\cite{Hoffberg} There is a critical neutron density ($1.45 \times
10 ^{17} Kg/ m^3$) below which s-wave pairing is dominant, as in
superconductors, but above which p-wave pairing takes over, as in
superfluid $^3 He$. Using the standard BCS relations, these
authors estimate that the transition temperature for the neutron
superfluid is $~10^{10} K$, well above the actual star
temperature.

Since the data collected from pulsars originates from the stars'
surface, atmosphere or magnetosphere, an important question
concerns whether differences in internal structure can be deduced
from observations of the pulsar pulses. Having said that, there
are phenomena that involve bulk dynamics and which should depend
on the internal composition These phenomena are typically the
radio pulsar "glitches" \cite{Lyne}, which are sudden increases in
the pulsar rotation rate often accompanied by an increase in slow
down rate followed by a period of relaxation (approximately
exponential, with time scale of days to years) towards the
pre-glitch frequency, and "timing noise", which consists of low
frequency structures.

During a glitch, the typical fractional increase in pulsar
rotation frequency is in the range $\Delta \nu/\nu = 10^{-9} \sim
10^{-6}$, and the relative increment in slow down rate $\Delta
\dot{\nu}/\dot{\nu}\sim10^{-3}$ where $\nu$ and $\dot{\nu}$ are
pulsar rotation frequency and frequency derivative respectively.
The trigger of the pulsar glitch is presently not well understood.
The long relaxation time associated with glitches is seen as
indirect evidence for neutron star superfluidity. On one side, in
the superfluid vortex unpinning and re-pinning model, triggering
of the glitch is due to coupling of the crust and the superfluid
interior as a consequence of a sudden unpinning of vortex lines
and the post-glitch relaxation is due to the vortex gradually
re-pinning to the crust lattice \cite{Anderson} \cite{Alpar},
\cite{Packard}. On the other side, in the classical starquake
model, as a consequence of the long-term spin-down in spin rate,
deformation stress in the rigid crust builds up to resist the
decreasing oblatness \cite{Baym1}. When the stress exceeds a
critical point, the crust cracks suddenly, resulting in a sudden
increase in spin rate. Based on the observed typical glitches,
both of the models have a sudden increase in rotation frequency
and slow down rate (i.e. $\Delta \dot{\nu}/\dot{\nu}$) at the time
of the glitch. The post-glitch relaxation represents a return to
equilibrium with a linear response of the interior superfluid,
while the lack of relaxation represents a non-linear response of
the superfluid \cite{Alpar}. As more glitches were detected, it
became clear that glitch behavior varies in aspects such as glitch
rate, amplitude and relaxation. Although, these diverse features
suggest glitches are triggered locally in the superfluid interior,
no model currently predicts the time between glitches or the size
of any given event (for two pulsars with similar rotation
parameters, one may glitch frequently while the other may never
have been observed to glitch).

As already mentioned above, in addition to glitches, pulsars also
suffer another kind of timing irregularity known as timing noise,
which is characterized by restless, unpredictable, smaller scale
fluctuations in spin rate \cite{Cordes1985} with time scales from
days to years. The timing noise induced fluctuations of pulse
frequency are small, with fractional change $\delta \nu/\nu <
10^{-9}$.Timing noise has been explained by random processes
\cite{Cordes1980}, unmodelled planetary companions or free
precession \cite{Stairs}. However the physical phenomenon
underlying most of the timing noise still has not been explained.

Presently the relation between glitches and timing noise is not
understood although Janssen and Stappers \cite{Janssen} showed
that it is possible to model the timing noise in PSR B1951+32 as
multiple small glitches. A better understanding of Pulsar timing
irregularities could lead to many important results, explaining
the cause of timing noise and glitches could allow us to relate
these phenomena and hence provide an insight into the interior
structure of neutron stars. However, currently it is still not
clear whether the glitch and timing noise phenomena are related.

\section{Glitches in Pulsars and Violation of the Weak Equivalence Principle for Superfluid Vortices}

Although the critical transition temperature of the neutron
superfluid, predicted by current physical models of neutron stars,
$10^8 K < T_c< 10^{10} K$, are many orders of magnitude above the
critical transition temperatures of ordinary superfluids in the
Earth laboratory, typically in the range of $1 K < T_c < 3 K$, we
will assume the possibility to extrapolate present superfluid
physics to the case of neutron superfluids in pulsars.

Since the major part of the mass of the star consists of neutrons,
it appears that most of the rotational energy resides in the
superfluid. By analogy with rotating He II, Ginzburg and Kirzhnits
\cite{Ginzburg}concluded that the neutron superfluid would contain
an array of quantised vortex lines. In the same way as in He II,
one can define critical values of angular velocity, $\Omega_{C1}$,
which must be exceeded to form a single vortex, and $\Omega_{C2}$,
at which the vortex cores would overlap \cite{Baym2} gave the
values $\Omega_{C1}\sim10^{-14} s^{-1}$ and
$\Omega_{C2}\sim10^{20} s^{-1}$; the periods of all known pulsars
correspond to rotational speeds ($\Omega$) ranging from over $1
s^{-1}$ to $10^3 s^{-1}$. Thus the neutron superfluid has
properties similar to He II undergoing solid-body rotation at the
temperature of $\sim 1 mK$. Thus we will apply the theoretical
model presented in section 3, which predicts that superfluid
vortices break the weak equivalence principle with an an
E\"{o}tv\"{o}s factor $\eta'$ given by eq.(\ref{v14}), to the case
of pulsars.

Substituting the following typical quantities in pulsars: for the
speed of sound in the neutron superfluid $c_s \sim c_0 10^{-6}
m/s$ \cite{Gusacov}, the length of a superfluid neutron vortex
$L\sim 1\times 10^3 m$, the radius of the vortex neutron core
$\xi=\hbar / m_{n^0} c_s\sim 2.1 \times 10^{-10} m$, the pulsar
angular velocity of $\Omega\sim 10^3 Rad/s$, the radius of the
superfluid neutron shell of $R=7,5\times 10^3 m$, and the neutron
density of $\rho=10^{17} Kg/m^3$ into eq.(\ref{v14}), one obtains
an E\"{o}tv\"{o}s factor $\eta'$
\begin{equation}
\eta' \sim 1.28 \times 10^{-8} \label{gp1}
\end{equation}
This quantifies the differential acceleration which would appear
between two neutron stars in free fall around another star, one
that would be rotating would host vortices, which would break the
WEP, and a second one that would not be rotating and would thus be
compliant with the WEP. Eq.(\ref{gp1}) also predicts the
differential angular velocity between two pulsars freely rotating
in the gravitomagnetic field generated by a third star, one which
do not break the WEP and a second one for which the superfluid
vortices break the WEP. Since neutron vortex lines would break the
WEP, a change in the effective number of vortex lines would cause
the moment of inertia of the fluid core, $I_f$, to change by the
fractional amount,
\begin{equation}
\frac{\Delta I_f}{I_f}=\frac {m_v}{\Delta m}\label{gp2}
\end{equation}
Using eq.(\ref{v14}), eq.(\ref{gp2}) can be expressed in function
of the E\"{o}tv\"{o}s factors $\eta$ and $\eta'$.
\begin{equation}
\frac{\Delta I_f}{I_f}=\frac{\eta'}{\eta}\label{gpp2}
\end{equation}
To conserve angular momentum the Pulsar crust must have a
fractional increase in angular speed given by:
\begin{equation}
\frac{\Delta \omega}{\omega}=\frac{\Delta I_f}{I_c}\label {gp3}
\end{equation}
where $I_c$ is the moment of inertia of the crust. Substituting
$\Delta I_f$ from eq.(\ref{gpp2}) in eq.(\ref{gp3}), we get:
\begin{equation}
\frac{\Delta\omega}{\omega}=\frac{I_f}{I_c}\frac{\eta'}{\eta}\label{gpp3}
\end{equation}
Since substituting the numerical values used above in this
section, in eq.(\ref{v10}) we obtain $\eta\sim1$, then
eq.(\ref{gp3}), simplifies to:
\begin{equation}
\frac{\Delta \omega}{\omega}\sim\frac{I_f}{I_c}\eta'\label {gp4}
\end{equation}
For any reasonable value of $I_f/I_c (10 - 100)$ \cite{Packard},
and substituting the value of $\eta'$ from eq.(\ref{gp1}) in
eq.(\ref{gp4}), the predicted speedups during glitches in pulsars,
$\Delta\omega/\omega (10^{-7}-10^{-6})$, would be in reasonable
agreement with currently observed values, which are in the typical
range $\Delta \nu/\nu (10^{-9} - 10^{-6})$. Thus we raise the
question of the possibility to apply our crude model for the
breaking of WEP for vortices in He II to the case of pulsars, as a
possible cause contributing to glitches and eventually other
timing noise in this type of stars.

Although the violation of the WEP in Pulsars within a predicted
E\"{o}tv\"{o}s factor $\eta'\sim 10^{-8}$, is not ruled out by
current observational tests of SEP in binary systems containing at
least one pulsar, (which, as referred above in section 2,  set an
upper limit on SEP violation with $|\eta|< 0.009$ at $95\%$
confidence level); the experimental detection of this phenomena in
these systems is a challenging task for current observational
capabilities, being 5 orders of magnitude away.

\section{Discussion}

In the Packard model \cite{Packard}, Glitches are accounted for by
various metastable states of a vortex array, which are possible
for a given angular velocity of the pulsar. In the neutron star, a
transition between two such states would involve a decrease in the
angular momentum of the superfluid, with a compensating increase
of of the angular momentum of the pulsar's crust. If the angular
momentum in the fluid core changes by the fractional amount
$\Delta L_f/ L_f$, then the crust must have a fractional increase
in speed of:
\begin{equation}
\frac{\Delta\omega}{\omega}=\frac{I_f}{I_c}\frac{\Delta
L_f}{L_f}\label{di1}
\end{equation}
comparing eq.(\ref{di1}) with eq.(\ref{gp4}), one concludes that
the Packard model is coherent with the breaking of WEP in pulsars
proposed in the present paper only if:
\begin{equation}
\eta'\sim\frac{\Delta L_f}{L_f}\label{di2}
\end{equation}
which means that both theoretical models depend on the effective
number of vortex lines in the neutron superfluid. However the
transfer of angular momentum from the rotating superfluid to the
pulsar crust, which is taking place in Packard model through
pinning and unpinning of vortex lines, is taking place in the
present model, through the change of the moment of inertia of the
superfluid core associated with a breaking of the WEP for neutron
vortex lines, according to eq.(\ref{gpp2}).

We wish to emphasize that the physical interpretation of glitches
as signs of a breaking of WEP in the pulsar neutron superfluid,
cannot provide by itself a physical mechanism capable to induce
transition from a pulsar state in which the WEP is broken to the
ground state which is compliant with the WEP.

The proposed model for the breaking of the WEP for vortex lines in
superfluid is also leaving us with a puzzle: Since the individual
atoms making the superfluid vortices (presumably) do satisfy the
weak equivalence principle, it is hard to see how vortices, as a
whole, can do other than obey weak equivalence. Therefore
attribution of large vortex mass excess to spontaneous gauge
symmetry breaking by Popov, Duan and others ought to imply either
that this symmetry breaking creates gravitational mass, or that it
breaks weak equivalence. In the present paper arguments have been
presented to support the latter physical possibility.

\section{Conclusions}

Since gauge invariance is broken in superconductors and
superfluids, it seems pertinent to investigate if the WEP is
broken for cooper pairs in superconductors and for vortices in
superfluids. Although, as pointed out by Cosimo Bambi in 2007
\cite{Bambi}, the interpretation of the anomalous Cooper pair
inertial mass excess in terms of a gravitomagnetic-type London
moment in rotating superconductors \cite{Tajmar02}, is not tenable
with respect to the experimental observation of orbital parameters
of pulsars in binary systems, as well as with respect to recent
experiments carried out by Tajmar et al. \cite{taj1}. It seems
that the physical interpretation of the anomalous mass of Cooper
pairs in superconductors in terms of a violation of the WEP for
this particles is more promising, since it is quite well accounted
for by an electromagnetic model of dark energy in the
superconductor \cite{de Matos1} \cite{matos}, and it leads to the
physical interpretation of the logarithmically diverging inertial
mass of vortices in rotating superfluids as a breaking of the WEP
for vortex lines \cite{de Matos2}. In the present paper we
demonstrated that the breaking of WEP for superfluid vortices,
predicted by our crude theoretical model, seems suitable to
account for glitches and timing noise in pulsars, at an
E\"{o}tv\"{o}s level $\eta'\sim 10^{-8}$. Although this prediction
is not ruled out by present tests of SEP in binary systems
containing at least one pulsar $|\eta|< 0.009$, it is
unfortunately too small, by 5 orders of magnitude, to be detected
by current astronomical observational capabilities.

\end{document}